\documentclass[twocolumn,english]{revtex4-1}
\usepackage[T1]{fontenc}
\usepackage[latin9]{inputenc}
\usepackage{color}
\usepackage{amsmath}
\usepackage{graphicx}
\usepackage{esint}

\makeatletter

\usepackage{babel}

\usepackage{babel}

\makeatother

\usepackage{babel}
\begin{document}
\title{Non-Fermi liquid behavior in the Sachdev-Ye-Kitaev model for a one
dimensional incoherent semimetal}
\author{Geo Jose$^{1}$, Kangjun Seo$^{2}$, Bruno Uchoa$^{1}$}
\affiliation{$^{1}$Center for Quantum Research and Technology, Department of Physics
and astronomy, University of Oklahoma, Norman, Oklahoma 73019, USA}
\affiliation{$^{2}$School of Electrical and Computer Engineering, The University
of Oklahoma, Tulsa, OK 74135, USA}
\date{\today}
\begin{abstract}
We study a two-band dispersive Sachdev-Ye-Kitaev (SYK) model in 1
+ 1 dimension. We suggest a model that describes a semimetal with
quadratic dispersion at half-filling. We compute the Green's function
at the saddle point using a combination of analytical and numerical
methods. Employing a scaling symmetry of the Schwinger-Dyson equations
that becomes transparent in the strongly dispersive limit, we show
that the exact solution of the problem yields a distinct type of non-Fermi
liquid with sublinear $\rho\propto T^{2/5}$ temperature dependence
of the resistivity. A scaling analysis indicates that this state corresponds
to the fixed point of the dispersive SYK model for a quadratic band
touching semimetal. 
\end{abstract}
\maketitle

\section{INTRODUCTION}

Sachdev-Ye-Kitaev (SYK$_{q}$) models describe strongly interacting
fermions with infinite range, $q-$body, random all-to-all interactions.
The $0+1$ dimensional SYK$_{q}$ dot model \cite{Sachdev,Sachdev2}
exhibits an approximate conformal symmetry in the infrared, is exactly
solvable in the limit of a large number of fermion flavors, saturates
the bound on quantum chaos and are dual to gravitational theories
in $1+1$ dimensions. \cite{Stanford,Stanford-1,Stanford-2}. Useful
connections to the black hole information problem have also been established
\cite{Stanford-3}.

In these models, approximate conformal symmetry in the strong coupling/low
frequency regime leads to power law behavior of correlation functions.
In the SYK$_{4}$ model, the zero temperature two-point correlation
function decays as $G\left(\omega\right)\sim1/\sqrt{\omega}$, with
$\omega$ the frequency. Finite temperature Green's functions can
be then obtained by appealing to conformal symmetry \cite{Sachdev}.
In dispersive versions of the SYK model, they result in the linear
temperature dependence of the dc resistivity, $\rho\propto T$, a
characteristic feature of strange metal phases. It was originally
conjectured that the linear scaling of the scattering rate in the
strange metal phase was due to hyperscaling in the proximity of a
quantum critical point buried inside the superconducting phase. Recent
momentum resolved electron energy spectroscopy experiments in the
cuprates revealed the emergence of a mysterious momentum independent
energy scale nearly one order of magnitude larger than the temperature
range of the quantum critical fan \cite{Mitrano,Husain}, at odds
with conventional quantum critical theories. One may speculate \cite{Zaanen,Patel3,Sachdev-1,Volovik,cha}
that the origin of the strange metal phase could be related to aspects
of the physics of incoherent metals.

Various lattice generalizations \cite{Zhang,Balents,Sachdev-1,Senthil,Shenoy,Ben-Zion}
of the dot model comprising of connected SYK dots have been recently
proposed in the regime where the SYK coupling is the dominant energy
scale of the problem. The general idea behind many of those extensions
is to build on the solution of the dot model including lattice effects
perturbatively. Weakly dispersive versions of the SYK model were used
to describe incoherent or `Planckian' metals which lack well defined
quasiparticles \cite{Patel2,Volovik,Dumitrescu}. These incoherent
metals typically have a crossover between the incoherent high temperature
regime and a low temperature Fermi liquid behavior\cite{Balents,Senthil}.
The crossover energy scale between the two regimes is set by $t^{2}/J$,
with $t$ proportional to the band width and $J\gg t$ the SYK coupling.
In the low temperature regime, the coherence of the quasiparticles
is restored by the presence of a large Fermi surface. Semimetals,
on the other hand, abridge a large class of gapless multiband systems
that lack a Fermi surface. One could ask what is the nature of the
normal state in a disordered semimetal with random local couplings.

\begin{figure}[b]
\begin{centering}
\includegraphics[scale=0.3]{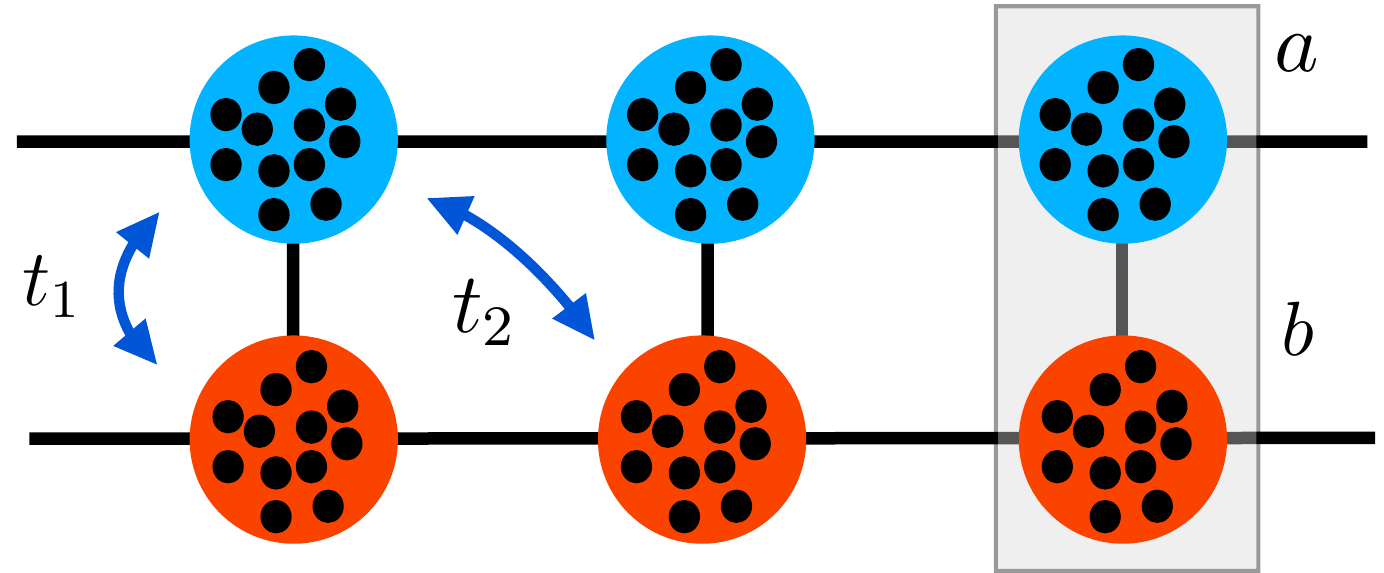}
\par\end{centering}
\caption{{\small{}Dispersive SYK ladder model: The unit cell contains two sites,
one for each chain (color). Each color site hosts $N$ complex fermions,
which interact locally through random couplings. We assume that hopping
is only allowed between different color sites, with $t_{1}$ the NN
hopping and $t_{2}$ the NNN one. The two-band quadratic dispersion
in Eq. \eqref{f} can be obtained by tuning $t_{1}=-2t_{2}=t$, with
$m=2/(ta^{2})$ the effective mass of the fermions, where $a$ is
the lattice constant.}}
\end{figure}

Motivated by these ideas, we study a 1D ladder with local random couplings
at every unit cell, as shown in Fig. 1. The hopping amplitudes between
lattice sites is finely tuned such that this system describes a half-filled
semimetal with quadratic dispersion and local SYK couplings. The weakly
dispersive limit $t^{2}/J\ll T\ll J$ has approximate conformal invariance
and recovers the usual SYK transport behavior, as expected. In the
strongly dispersive regime, $T\ll t^{2}/J$, the scaling symmetry
of the problem becomes transparent, albeit the absence of conformal
symmetry. In this limit, the incoherent regime extends down to zero
frequency and temperature, unlike in the more conventional metallic
case. We show that the resistivity of this model has a sublinear scaling
with temperature, 
\begin{equation}
\rho\propto T^{2/5},\label{rho}
\end{equation}
whereas the Lorentz ratio $L=\kappa/(\sigma T)\approx3.2$ is fairly
close to the value expected for a Fermi liquid, $L=\pi^{2}/3$. We
find through a scaling argument that when the system starts from the
SYK fixed point at high temperature, it flows towards a distinct non-Fermi
liquid (NFL) fixed point at zero temperature. At intermediate energy
scales, away from the low temperature fixed point, the system crosses
over from a ``Planckian'' semimetal to an incoherent NFL with sublinear
temperature scaling of the resistivity.

This paper is organized in the following way: in section II we introduce
the lattice model of a 1D ladder of SYK quantum dots that behaves
as a 1D semimetal with quadratic dispersion. In section III we address
the Green's function of this system at zero and finite temperature.
In the strongly dispersive regime ($T\ll t^{2}/J$), where conformal
symmetry is not present, we numerically extract the finite temperature
scaling functions of the Green's function and of the self-energy.
In section IV, we discuss a scaling analysis of the problem and the
crossover between the high temperature incoherent Planckian regime
and the low temperature NFL one. In section V we address the temperature
scaling of the dc resistivity of the model. Finally, in section VI
we present our conclusions.

\section{MODEL}

\noindent We consider $N-$flavors of complex fermions hopping on
an 1D lattice. Each lattice site hosts an SYK dot with random, site
dependent interactions $J_{ijk\ell}^{x}$ between them. The indexes
$i,j,k,\ell=1,..,N$ label the $N-$flavors/colors per site. We start
from a 1D ladder shown in Fig. 1, with two sites per unit cell, shown
in blue and red. Allowing hopping processes between blue and red sites
only, the Hamiltonian of the kinetic term can be written as
\begin{equation}
\hat{\mathcal{H}}_{0}=\int_{k}\sum_{\nu}f(k)\psi_{k,i,\nu}^{\dagger}\left(\sigma_{x}\right)_{\nu,\nu^{\prime}}\psi_{k,i,\nu^{\prime}},\label{eq:H01}
\end{equation}
where
\begin{equation}
f(k)=t_{1}+2t_{2}\cos(ka),\label{f}
\end{equation}
with $t_{1}$ and $t_{2}$ the hopping between nearest neighbors (NN)
and second nearest neighbors (NNN) respectively among different color
sites, and $\int_{k}\equiv a(2\pi)^{-1}\int_{-\Lambda}^{\Lambda}\text{d}k$
with $\Lambda=\pi/a$ the ultraviolet cutoff. $\psi_{i,\nu}$ is a
two-component spinor in the site basis of the unit cell $\nu=a,b$
and $\sigma_{x}$ is the real off-diagonal Pauli matrix in that basis.
Fine tuning the hopping constants to $t_{1}=-2t_{2}=t$, then
\begin{equation}
\mathcal{H}_{0}=\int_{k}\sum_{i\nu}\frac{k^{2}}{2m}\psi_{k,i,\nu}^{\dagger}\left(\sigma_{x}\right)_{\nu\nu^{\prime}}\psi_{k,i,\nu^{\prime}},\label{H02}
\end{equation}
in the continuum limit ($k\ll1/a$), with $m^{-1}=ta^{2}/2$. The
dispersion of this model has two quadratic bands touching at a single
point $k=0$, as shown in Fig. 2. At half filling, the band lacks
a Fermi momentum and behaves as a 1D semimetal. In the following,
we assume the band to be half filled and set $a\to1$. 
\begin{figure}
\centering{}\includegraphics[scale=0.3]{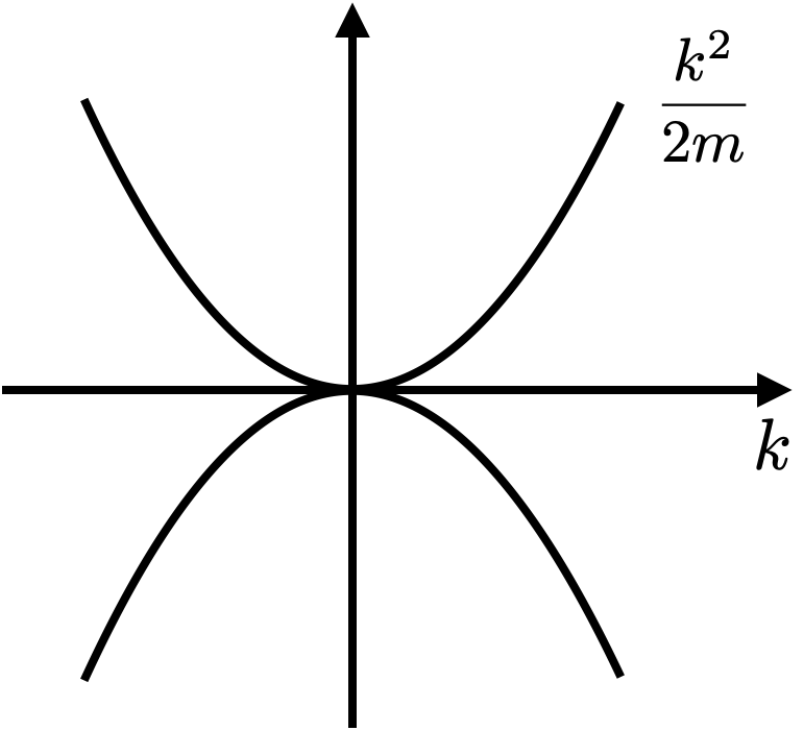}\caption{{\small{}Finely tuned energy dispersion for the ladder model illustrated
in Fig. 1. The half-filled band describes a 1D semimetal with parabolic
band touching at $k=0$. }}
\end{figure}

These fermions interact via a local, instantaneous two body SYK interaction,
\begin{equation}
\mathcal{H}_{\text{SYK}}=\frac{1}{\left(N\right)^{\frac{3}{2}}}\!\!\sum_{\nu\nu',ijk\ell}\int_{x}J_{ijk\ell}^{x}\psi_{x,i,\nu}^{\dagger}\psi_{x,j,\nu^{\prime}}^{\dagger}\psi_{x,k,\nu'}\psi_{x,\ell,\nu}\label{HI}
\end{equation}
with random, color site independent matrix elements $J_{ijk\ell}^{x}$
that are properly antisymmetrized with $J_{ijk\ell}^{x}=-J_{jik\ell}^{x}=-J_{ij\ell k}^{x}.$
As in the other SYK models, we take these to be complex Gaussian distributed
coupling with a zero mean value $\langle\langle J_{ijk\ell}^{x}\rangle\rangle=0$
and variance $\langle\langle|J_{ijk\ell}^{x}|^{2}\rangle\rangle=J^{2}/8.$

The standard method to study the current problem is the imaginary
time path integral formalism, where the partition function is given
by $\mathcal{Z}=\int\left[\mathcal{D}\bar{\psi}\mathcal{D}\psi\right]\mathrm{e}^{-\mathcal{S}}$,
where $\mathcal{S}=\mathcal{S}_{0}+\mathcal{S}_{\text{SYK}}$, with
\begin{equation}
\mathcal{S}_{0}=\int_{\tau,x}\sum_{\ell,\nu}\bar{\psi}_{\ell,\nu}\left(\tau,x\right)\left[\partial_{\tau}-\left(\sigma_{x}\right)_{\nu\nu^{\prime}}\frac{\partial_{x}^{2}}{2m}\right]\psi_{\ell,\nu^{\prime}}\left(\tau,x\right).\label{So}
\end{equation}
We define $\int_{\tau}\equiv\int_{0}^{\beta}\text{d}\tau$, with $\beta=1/T$,
and $\mathcal{S}_{\text{SYK}}$ the corresponding two body action
of (\ref{HI}) with with same time Grassmann fields $\bar{\psi}(\tau,x)$
and $\psi(\tau,x)$. In order to deal with the disorder, we use the
replica trick and average over disorder realizations. This procedure
amounts to an annealed approximation\cite{Rosenhaus}.\textcolor{blue}{{}
}Using
\begin{equation}
\langle\langle e^{-\sum_{ijk\ell}J_{ijk\ell}A_{ijk\ell}}\rangle\rangle=e^{J^{2}\sum_{ijk\ell}\bar{A}_{ijk\ell}A_{ijk\ell}}\label{average}
\end{equation}
and defining the Green's function 
\begin{equation}
\hat{G}_{\nu\nu^{\prime}}\left(\tau,x\right)=-\frac{1}{N}\sum_{\ell=1}^{N}\langle T[\psi_{\nu,\ell}\left(0,0\right)\bar{\psi}_{\nu^{\prime},\ell}\left(x,\tau\right)]\rangle,\label{G0-1}
\end{equation}
the integration over the fermionic fields results in the saddle point
action, \begin{widetext}
\begin{multline}
\mathcal{S}_{\text{eff}}=-\log\det\left[\delta\left(\tau-\tau'\right)\delta\left(x-x'\right)\left(\partial_{\tau}+\sigma_{x}\left(i\partial_{x}\right)^{2}\right)+\hat{\Sigma}\left(x-x',\tau-\tau'\right)\right]\\
-\frac{J^{2}}{8}\int_{\tau,\tau^{\prime}}\text{tr}\hat{G}^{2}\left(0,\tau-\tau'\right)\text{tr}\hat{G}^{2}\left(0,\tau'-\tau\right)-\int_{x,x^{\prime}}\int_{\tau,\tau^{\prime}}\text{tr}\left[\hat{\Sigma}\left(x-x',\tau-\tau'\right)\hat{G}\left(x'-x,\tau'-\tau\right)\right],\label{Seff}
\end{multline}
\end{widetext}where $\hat{\Sigma}\left(x-x',\tau-\tau'\right)$ is
the self-energy. The action can be minimized exactly in $\hat{G}$
and $\hat{\Sigma}$ in the large-$N$ limit. Following the minimization,
the solutions form a set of Schwinger-Dyson equations
\begin{equation}
\hat{G}^{-1}\left(i\omega_{n},k\right)=i\omega_{n}-\frac{k^{2}}{2m}\sigma_{x}-\hat{\Sigma}\left(i\omega_{n},k\right),\label{G}
\end{equation}
and
\begin{equation}
\hat{\Sigma}\left(\tau,x\right)=-\frac{J^{2}}{2}\delta(x)\hat{G}\left(-\tau,0\right)\mathrm{tr}\!\left[\hat{G}\left(\tau,0\right)\hat{G}\left(\tau,0\right)\right],\label{Sigma}
\end{equation}
The self-energy $\hat{\Sigma}\left(i\omega_{n},k\right)\equiv\hat{\Sigma}\left(i\omega_{n}\right)$
is therefore momentum independent, reflecting the $x-$dependence
of the couplings $J_{ijk\ell}^{x}$. The disorder averaged SYK term
is uncorrelated and purely local. We denote the Fourier transform
of the momentum independent self-energy as $\hat{\Sigma}\left(\tau\right)$.
We also denote
\begin{equation}
\hat{\mathcal{G}}(\tau)\equiv\hat{G}(\tau,0)=\int_{k}\hat{G}(\tau,k)\label{eq:G}
\end{equation}
for the momentum integrated Green's function.

\begin{figure*}
\includegraphics[scale=0.43]{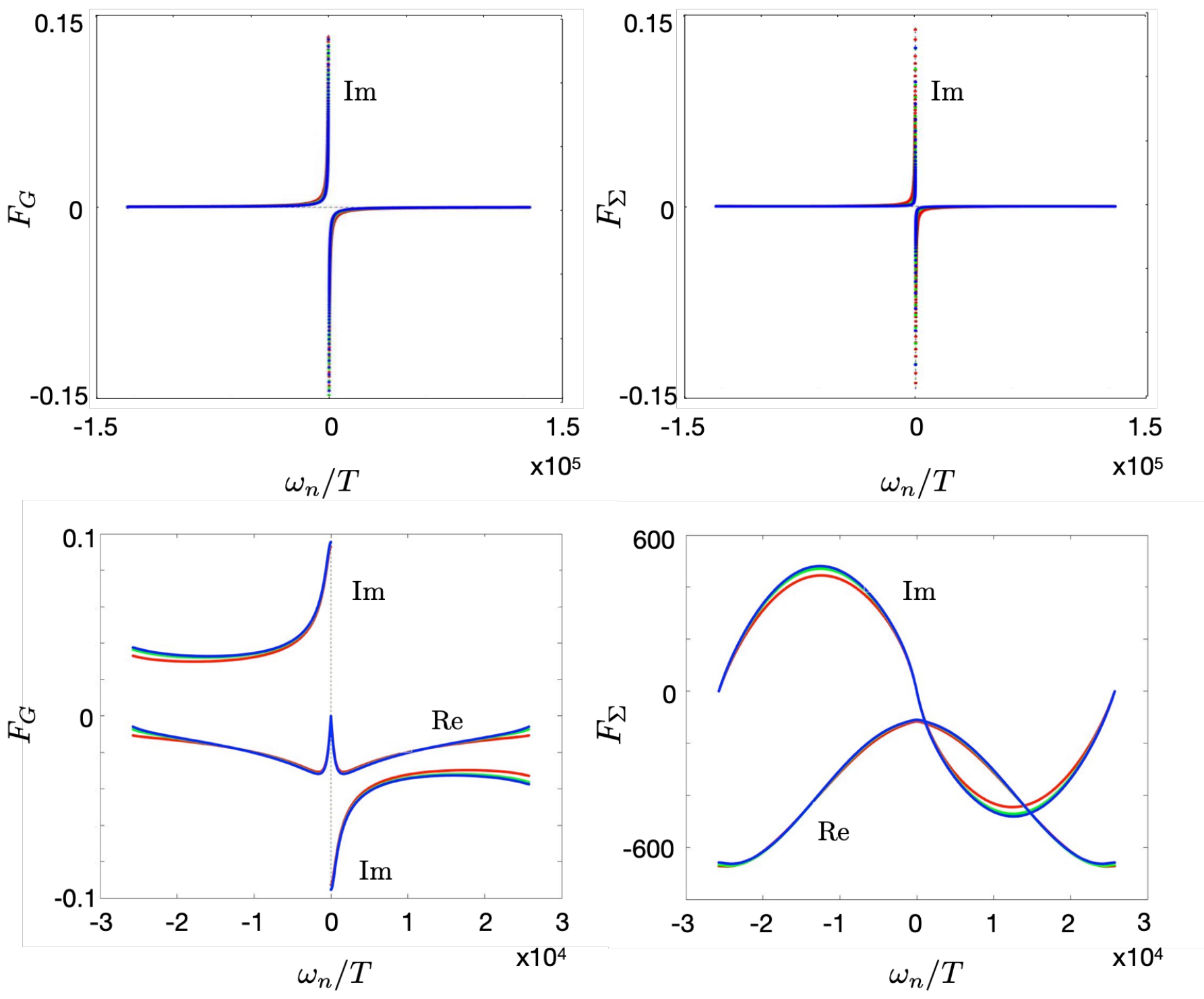}\caption{{\small{}Scaling functions for the Green's function ($F_{G}$) and
the self-energy ($F_{\Sigma}$) versus Matsubara frequency $\omega_{n}$
normalized by temperature $T$. Top row: numerical solution of Eq.
(\ref{SD2}) and (\ref{SD3}) for $F_{G}$ and $F_{\Sigma}$ in the
SYK limit for various temperatures, namely $T=10,\,20$ and $30$
(green line, blue and red, respectively). $\beta=T^{-1}$and $J=100$
are set in units of $2m$ with $a\to1$ ($\Lambda=\pi$). In this
case, the Green's function and self energy are purely imaginary and
admit an analytical solution {[}see Eq. (\ref{eq:Sigma-1}){]}. Bottom
row: numerical solution of Eqs. (14) and (25), in the strongly dispersive
regime. The real and imaginary parts of the scaling functions were
computed at various temperatures, namely $\beta=1/T=256,\,64$ and
$4$ (green line, blue and red, respectively). All curves nearly coincide
at low frequencies, where the scaling functions are expected to be
temperature independent. }}
\end{figure*}

\section{GREEN'S FUNCTION}

If one takes the ansatz for the Green's function, $\hat{G}(\tau)=\mathcal{G}(\tau)\sigma_{x}$,
the self energy then has to be of the form $\hat{\Sigma}(\tau)=\Sigma(\tau)\sigma_{x}$.
As in usual SYK models, we make the usual infrared assumption $i\omega_{n}\ll\Sigma(i\omega_{n})$.
The Schwinger Dyson equations (\ref{G}) and (\ref{Sigma}) can be
written as 
\begin{align}
\mathcal{G}(i\omega_{n}) & =-\int_{k}\frac{1}{\frac{k^{2}}{2m}+\Sigma(i\omega_{n})}\nonumber \\
 & =-\frac{\sqrt{2m}}{\pi\sqrt{\Sigma(i\omega)}}\text{tan}^{-1}\left(\frac{\Lambda}{\sqrt{2m}\sqrt{\Sigma(i\omega)}}\right),\label{SD1}
\end{align}
and 
\begin{equation}
\Sigma(\tau)=-J^{2}\mathcal{G}^{2}(\tau)\mathcal{G}(-\tau),\label{SD2}
\end{equation}
There are two limits of particular interest. As it will be clear in
the next section, one is the intermediate frequency limit $t^{2}/J\ll\omega\ll J$,
where the argument of the $\tan^{-1}(y)$ function, 
\begin{equation}
y\equiv\frac{\Lambda}{\sqrt{2m}\sqrt{\Sigma(i\omega)}}\label{eq:y}
\end{equation}
is small. This regime corresponds to the weakly dispersive limit,
which recovers the physics of the $0+1$ dimensional SYK dot. The
other is the strongly dispersive regime, $\omega\ll t^{2}/J$, where
$y\gg1$. We show that this limit is exactly solvable and leads to
a different NFL regime.

\subsection{Weakly dispersive limit}

In this weakly dispersive limit ($y\ll1)$, the SYK physics dominates
and typically one obtains a fully incoherent system with a linear
in $T$ dc resistivity. In that regime, Eq. (\ref{SD1}) becomes
\begin{equation}
\mathcal{G}(i\omega_{n})\approx-\frac{1}{\pi}\frac{\Lambda}{\Sigma\left(i\omega_{n}\right)}.\label{SD3}
\end{equation}
The Schwinger-Dyson Eq. (\ref{SD2}) and (\ref{SD3}) have the same
form as in the SYK dot model \cite{Sachdev}. They have conformal/reparametrization
invariance, indicating a power law solution at $T=0$.

At zero temperature, Eq. (\ref{SD2}) and (\ref{SD3}) can be solved
by the ansatz $\mathcal{G}\left(i\omega\right)=c_{1}\text{e}^{-i\frac{\pi}{4}}\left(i\omega\right)^{-\frac{1}{2}}$
for the time ordered Green's function \cite{Sachdev}. Using this
result in Eq. (\ref{SD2}) and taking a Fourier transform one finds
\begin{equation}
\Sigma\left(i\omega\right)=-\frac{J^{2}c_{1}^{3}}{\pi}e^{i\frac{\pi}{4}}\,\sqrt{i\omega},\label{eq:Sigma-1}
\end{equation}
where the constant $c_{1}=\Lambda^{\frac{1}{4}}/\sqrt{J}.$ The zero
temperature dispersive Green's function is
\begin{equation}
G(i\omega,k)=\frac{1}{\Sigma\left(i\omega\right)}-\frac{k^{2}/2m}{\Sigma\left(i\omega\right)^{2}}+\ldots\,.\label{eq:G-1}
\end{equation}
This solution introduces a perturbative correction to the SYK Green's
function, in the same spirit as in the metallic case \cite{Senthil}.

To get the finite temperature solutions, one can then use the conformal
map, $\tau\rightarrow f\left(\tau\right)=\tan\frac{\pi\tau}{\beta}$.
Applying this to the Fourier transform of $\mathcal{G}\left(i\omega\right)$
gives
\begin{equation}
\mathcal{G}\left(\tau\right)=\mathrm{sgn}(\tau)\,c_{1}\sqrt{\frac{1/\beta}{2\sin\pi\tau/\beta}}.\label{Gtau}
\end{equation}
The finite temperature self-energy $\Sigma\left(i\omega_{n}\right)$
can then be obtained from a Fourier transform of (\ref{SD2}),
\begin{equation}
\Sigma\left(i\omega_{n}\right)=iJ^{2}c_{1}^{3}\frac{\left(2\pi\right)^{3/2}\sqrt{2}\Gamma\left(\frac{3}{4}+\frac{\omega_{n}\beta}{2\pi}\right)\Gamma\left(-\frac{1}{2}\right)}{\sqrt{\beta}\pi\Gamma\left(\frac{1}{4}+\frac{\omega_{n}\beta}{2\pi}\right)},\label{omega}
\end{equation}
with $\omega_{n}$ a Matsubara frequency. The dispersive Green's function
at finite temperature follows from Eq. (\ref{SD3}) and (\ref{omega}),
\begin{equation}
G(i\omega_{n},k)=\frac{1}{\Sigma\left(i\omega_{n}\right)}-\frac{k^{2}/2m}{\Sigma^{2}\left(i\omega_{n}\right)}+\ldots\,.\label{G5}
\end{equation}

It is well known that the Greens function and self energies in this
regime have some convenient scaling properties. Eqs. (\ref{SD2})
and (\ref{SD3}) admit solutions of the form:

\begin{gather}
\mathcal{G}(i\omega_{n})=(JT)^{-\frac{1}{2}}\Lambda^{-\frac{3}{4}}F_{G}\left(\frac{\omega_{n}}{T}\right)\label{Gomega2-1}\\
\Sigma(i\omega_{n})=(JT)^{\frac{1}{2}}\Lambda^{\frac{3}{4}}F_{\Sigma}\left(\frac{\omega_{n}}{T}\right)\label{scalingform-1}
\end{gather}
where $F_{G,\Sigma}$ are scaling functions which are independent
of all parameters. The scaling functions are plotted in the top row
panels of Fig. 3. They were obtained by numerically solving Eq. (\ref{SD2})
and (\ref{SD3}) for various temperatures and then stripping away
the power law dependence in $T$ and $J$ from the above equations.
The results show good agreement with the scaling arguments.

\subsection{Strongly dispersive regime}

Next we consider the regime where $y\gg1$. As we will show below,
this inequality corresponds to the regime where
\begin{equation}
\omega\ll\frac{t^{2}}{J},\label{iomega t}
\end{equation}
and leads a different kind of NFL behavior compared to weakly dispersive
SYK models.

In the $y\gg1$ regime, the Schwinger-Dyson equation (\ref{SD1})
reads
\begin{equation}
\mathcal{G}(i\omega)=-\frac{1}{2}\frac{\sqrt{2m}}{\sqrt{\Sigma(i\omega)}}.\label{SD4}
\end{equation}
Eq. \eqref{SD2} and \eqref{SD4} admit a power law solution at $T=0,$
given by the ansatz 
\begin{equation}
\mathcal{G}(\tau)=C\frac{1}{\left|\tau\right|^{2\Delta}},\label{eq:Git}
\end{equation}
where $2\Delta=3/5$, as found in a related model \cite{Shenoy},
with
\begin{equation}
C=\left[\frac{\Gamma(\frac{1}{5})^{-1}\Gamma(\frac{2}{5})^{-2}}{20\sin^{2}\left(\frac{3\pi}{10}\right)\sin\left(\frac{9\pi}{10}\right)}\right]^{\frac{1}{5}}\approx0.40.\label{C}
\end{equation}
That solution corresponds to a self-energy
\begin{equation}
\Sigma(i\omega)=C^{\prime}J^{\frac{4}{5}}m^{\frac{3}{5}}\left|\omega\right|^{6\Delta-1}\gg|\omega|,\label{sigmaomega}
\end{equation}
from equation (\ref{SD2}). Explicit verification of this solution
follows by Fourier transforming (25),
\begin{equation}
\mathcal{G}(i\omega)=2C\sin\pi\Delta\ \Gamma(1-2\Delta)J^{-\frac{2}{5}}m^{\frac{1}{5}}\left|\omega\right|^{2\Delta-1},\label{gtau}
\end{equation}
and calculating $\Sigma\left(i\omega\right)$ from (\ref{SD2}). The
zero-temperature Green's function is hence
\begin{equation}
\hat{G}(i\omega,k)=\frac{-\sigma_{x}}{\frac{k^{2}}{2m}+C^{'}J^{\frac{4}{5}}m^{\frac{3}{5}}\left|\omega\right|^{4/5}},\label{G-1}
\end{equation}
where 
\begin{equation}
C^{\prime}=-2C^{3}\sin\left(\frac{9\pi}{10}\right)\Gamma\left(-\frac{4}{5}\right)\approx0.22.\label{C'}
\end{equation}
The Green's function above describes a distinct type of incoherent
semimetal, and is valid all the way down to zero frequency. It contrasts
with the result in the coherent $J\to0$ limit of the ladder problem
model ($J/|\omega|\ll1$), where the Green's function has a pole with
well defined quasiparticles. One needs to analytically continue the
above solution and impose physical restrictions to obtain the exact
Green's function. 

Note that the linear in $T$ resistivity in SYK models stems from
$\Delta=\frac{1}{4}.$ In strongly dispersive semimetals, with $y\gg1$,
finite temperature solutions cannot be obtained using a conformal
map on the $T=0$ solution because \eqref{SD2} and \eqref{SD4} do
not have the requisite conformal/reparametrization symmetry. Finite
temperature solutions to these equations then have to be obtained
numerically. However, we still have a scaling symmetry which dictates
a certain scaling form for the solutions.

Rewriting \eqref{SD4} in $\tau-$space, 
\begin{equation}
\int_{\tau_{1},\tau_{2}}\mathcal{G}(\tau)\mathcal{G}(\tau-\tau_{2})\Sigma(\tau-\tau_{1})=\frac{m}{2}\delta(\tau).
\end{equation}
It is easy to see that these equations are invariant under
\begin{equation}
\tau\rightarrow b\tau,\quad\mathcal{G}\rightarrow b^{-\frac{3}{5}}\mathcal{G},\quad\Sigma\rightarrow b^{-\frac{9}{5}}\Sigma.\label{scaling}
\end{equation}
Under this scaling, $T\rightarrow T/b$ leaving $T\tau$ invariant.
With this information, one can see that \eqref{SD2} and \eqref{SD4}
admit a solution of the form
\begin{equation}
\mathcal{G}(\tau)=m^{\frac{1}{5}}J^{-\frac{2}{5}}T^{\frac{3}{5}}\tilde{\mathcal{G}}(T\tau)\label{gtau2}
\end{equation}
 and
\begin{equation}
\Sigma(\tau)=m^{\frac{3}{5}}J^{\frac{4}{5}}T^{\frac{9}{5}}\tilde{\Sigma}(T\tau).\label{sigmatau2}
\end{equation}

Equivalently in Fourier space, we find
\begin{gather}
\mathcal{G}(i\omega_{n})=m^{\frac{1}{5}}(JT)^{-\frac{2}{5}}F_{G}\left(\frac{\omega_{n}}{T}\right)\label{Gomega2}\\
\Sigma(i\omega_{n})=m^{\frac{3}{5}}(JT)^{\frac{4}{5}}F_{\Sigma}\left(\frac{\omega_{n}}{T}\right)\label{scalingform}
\end{gather}
where $F_{G,\Sigma}$ are scaling functions that are independent of
temperature $T$ and the coupling $J$, with $\omega_{n}$ a Matsubara
frequency. The dispersive finite temperature Green's function of the
problem in this regime is 
\begin{equation}
\hat{G}^{-1}(i\omega_{n},k)=-\left(\frac{k^{2}}{2m}+m^{\frac{3}{5}}(JT)^{\frac{4}{5}}F_{\Sigma}(\omega_{n}/T)\right)\sigma_{x}.\label{g5}
\end{equation}
As shown in the next section, this will suffice to determine the temperature
dependence of various transport coefficients. Strictly speaking, the
scaling symmetry is only present in the infrared limit of the theory.
This means that the exact numerical solutions may violate these scaling
forms at very high frequencies. The real and imaginary parts of the
numerically obtained scaling functions $F_{G}$ and $F_{\Sigma}$
are plotted in the bottom row of Fig. 3. The plots show good agreement
with Eq. \eqref{scalingform} even outside the infrared limit.

\section{Scaling analysis}

After averaging over the disorder, which is spatially uncorrelated,
the SYK term in the action has eight fermionic fields, which we symbolically
write as 
\begin{equation}
\mathcal{S}_{\text{SYK}}=J^{2}\int_{\tau_{1},\tau_{2},x}[\bar{\psi}(\tau_{1},x)\psi(\tau_{1},x)]^{2}[\bar{\psi}(\tau_{2},x)\psi(\tau_{2},x)]^{2}.\label{SSYK2}
\end{equation}
Rescaling time as $\tau^{\prime}=\tau/s$ and imposing the SYK coupling
$J$ to be marginal, then the fields rescale as $\psi^{\prime}=\psi/s^{\frac{1}{4}}$.
As pointed out before \cite{Balents,Senthil}, analyzing the problem
in the vicinity of the fixed point of the $0+1$ dimensional SYK model
$(t=0$), the kinetic term is a relevant perturbation and the hopping
parameter grows as $t^{\prime}=t\sqrt{s}$. If one starts from temperature
$T$ in the weakly dispersive regime $t\ll J$, the hopping will grow
until $t^{\prime}(s_{*})\sim J$ for a rescaling parameter no larger
than $s_{*}=J/T$. Hence, the scaling stops at $T=t^{2}/J$. The validity
of the incoherent ``Planckian'' regime requires that
\begin{equation}
T\gg T_{*}=t^{2}/J,\label{T*}
\end{equation}
as in the case of incoherent metals with a finite Fermi surface \cite{Senthil}.

If one continues to lower the temperature further below $T_{*}$,
we claim that the system crosses over to a different type of incoherent
NFL regime with sublinear temperature scaling of the resistivity,
as illustrated in Fig. 4. From the perspective of the Schwinger-Dyson
equations \eqref{SD1} and \eqref{SD2}, the parameter that controls
the crossover between the weakly and the strongly dispersive regimes
is
\begin{equation}
y(T)\sim\sqrt{\frac{t}{\Sigma(T)}}.\label{y2}
\end{equation}
In the strongly dispersive regime $y(T)\gg1$, setting $m\sim t^{-1}$,
one can write the solution of the finite temperature self-energy \eqref{scalingform}
as
\begin{equation}
\Sigma(T)\propto t^{-\frac{3}{5}}(JT)^{\frac{4}{5}}\ll t.\label{SIGMA3}
\end{equation}
This inequality leads to $T\ll T_{*}=t^{2}/J$. In the same way, in
the weakly dispersive regime ($y(T)\ll1$),
\begin{equation}
\Sigma(T)\propto\sqrt{JT}\gg t,\label{SIGMA4}
\end{equation}
 as seen from Eq. \eqref{omega}, implying that $T\gg T_{*}$.

We note that in the case of metals, the $T<T^{*}$ regime was found
to realize a Fermi liquid. In the case of a 1D half-filled semimetal
with parabolic touching, we showed that the low temperature regime
does not lead to a semimetal, but to another type of incoherent NFL,
whose transport properties will be addressed in the next section.

\begin{figure}
\includegraphics[scale=0.4]{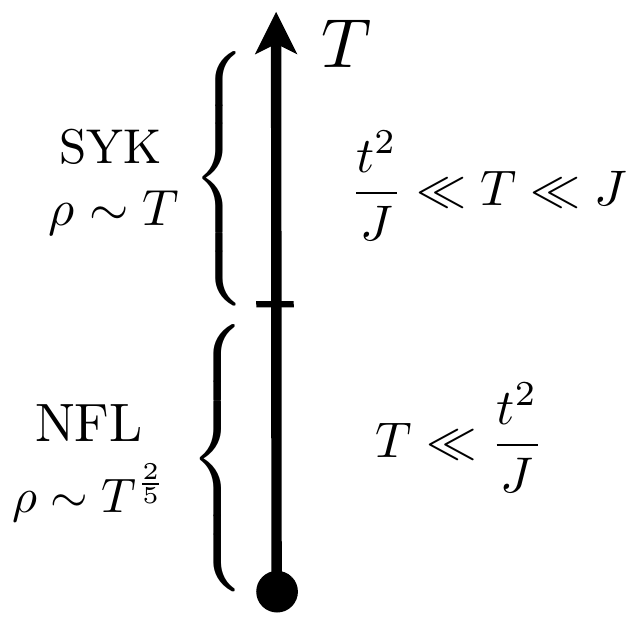}

\caption{{\small{}Different temperature regimes in the scaling. At temperature
$T>T_{*}=t^{2}/J$, the system is close to the $0+1$ dimensional
SYK fixed point and shows Planckian behavior, with linear dependence
of the resistivity in temperature. Below $T_{*}=t^{2}/J$ the system
crosses over towards a distinct type of incoherent NFL, with $\rho\sim T^{\frac{2}{5}}$.}}
\end{figure}

\section{TRANSPORT}

In this section, we look at the electric and thermal conductivities
using the above finite temperature solutions. These can be computed
using the Kubo formula. The charge current operator for this model
is 
\begin{equation}
\hat{j}_{e}=\frac{e}{2m}\int_{k}\sum_{i}k\psi_{ki}^{\dagger}\sigma_{x}\psi_{ki}.\label{j}
\end{equation}
The zero frequency conductivity is given by 
\begin{equation}
\sigma_{\text{dc}}=\mathrm{lim}_{\omega\to0}\frac{\mathrm{Im}\mathrm{\mathcal{K}}^{\text{ret}}(\omega)}{\omega},\label{sigmadc}
\end{equation}
where $\mathrm{\mathcal{K}}^{\text{ret}}(\omega)$ is the retarded
current-current correlation function, $\mathrm{\mathcal{K}}(\tau)=\langle T[\hat{j}_{e}(0,x)\hat{j}_{e}(\tau,x)\rangle$,
given by the series of diagrams in Fig. 5.\textbf{\textcolor{red}{{}
}}Each diagram displayed in that figure is of order $N$. For instance,
in diagram (b) each SYK vertex contributes a factor of $N^{-\frac{3}{2}}$
, while the three independent flavor sums contribute $N^{4}$, making
it a total of order $N.$ Diagrams in (b) and (c) vanish because the
current vertex is an odd function of momenta. This can be readly seen
by noticing that because of the disorder averaging, the summation
over momenta through each current vertex can be performed independently,
resulting in a zero contribution of those diagrams \cite{Georges}.
The remaining diagram is shown in Fig. 5a. It can be written in terms
of the Green's functions derived before as

\begin{equation}
\mathcal{K}(i\omega_{n})=\frac{Ne^{2}}{(2m)^{2}}T\ \mathrm{tr}\sum_{\nu_{n}}\int_{k}k^{2}\hat{G}(i\nu_{n},k)\hat{G}(i\nu_{n}+i\omega_{n},k).\label{eq:Kappa_E}
\end{equation}
The transport properties of the weakly dispersive regime recovers
the expected behavior of incoherent metals found in Ref. \cite{Sachdev-1,Senthil},
$\sigma_{\text{dc}}\propto1/T$, and we will focus instead in the
strongly dispersive case.

It is usually challenging to sum over Matsubara frequencies in the
absence of poles in the Green's functions. One can circumvent that
difficulty by using the spectral representation of the Green's function
\begin{equation}
\hat{G}(i\omega_{n},k)=\int_{\omega'}\frac{\mathcal{\hat{A}}(k,\omega')}{i\omega_{n}-\omega'},\label{eq:Gspect}
\end{equation}
with
\begin{equation}
\mathcal{\hat{A}}(k,\omega)=\frac{1}{\pi}\sigma_{x}\mathrm{Im}\frac{1}{\frac{k^{2}}{2m}+\Sigma(\omega+i0_{+})}\label{A}
\end{equation}
 the spectral function. One arrives at
\begin{equation}
\sigma_{dc}=\frac{Ne^{2}\sqrt{2m}}{\pi^{2}T}\int_{\omega}\frac{\left[\text{Im}\Sigma(\frac{\omega}{T})\right]^{2}}{\cosh^{2}(\frac{\omega}{2T})}\int_{k}\frac{k^{2}}{\left|\frac{k^{2}}{2m}+\Sigma\left(\frac{\omega}{2T}\right)\right|^{4}}.\label{sigmadc-1}
\end{equation}
Equivalently, casting Eq. \eqref{sigmadc-1} in terms of the scaling
functions \eqref{scalingform}, the dc conductivity is 
\begin{equation}
\sigma_{\text{dc}}(T)=\frac{Ne^{2}}{\sqrt{2m}(JT)^{\frac{2}{5}}}I_{1},
\end{equation}
where
\begin{equation}
I_{1}=\frac{1}{\pi^{2}}\int_{0}^{\infty}\text{d}z\frac{\left[\text{Im}F_{\Sigma}(z)\right]^{2}}{\cosh^{2}(\frac{z}{2})}\int_{0}^{\infty}\text{d}y\frac{y^{2}}{\left|y^{2}+F_{\Sigma}\left(z\right)\right|^{4}}\label{I1}
\end{equation}
is a dimensionless integral, and $F_{\Sigma}(z)$ the analytically
continued scaling function of the self energy.

\begin{figure}
\includegraphics[scale=0.3]{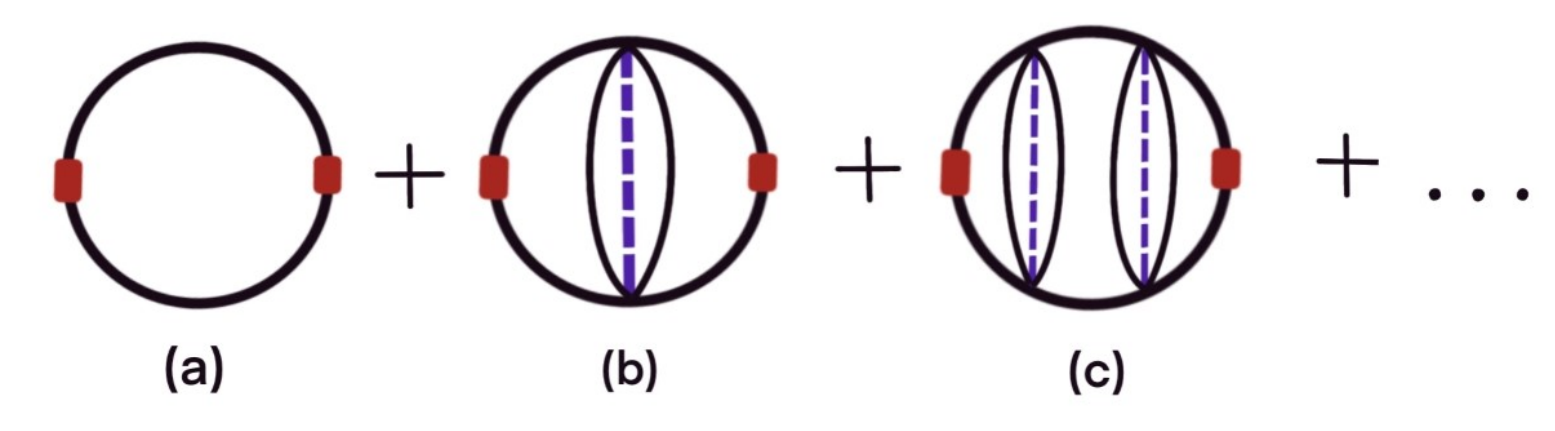}

\caption{Diagrams that contribute to the current-current correlation function
to leading order in $N$ (see text). Red rectangles represent the
current vertex. Black lines represent the fermion Green\textquoteright s
function and dotted blue line represents disorder average. Since the
current vertex is an odd function of momenta, diagrams (b) and (c)
and so on vanish, leaving (a) as the sole contribution in the large-$N$
limit. }
\end{figure}

A signature property of Fermi liquids is captured by the Lorentz ratio,
$L=\kappa/(\sigma T)$, which is the ratio of thermal $(\kappa)$
and electric $(\sigma)$ conductivities. In the particle-hole symmetric
model, the thermoelectric contribution to the thermal conductivity,
which is proportional to $T/\sigma$, vanishes by symmetry and can
be ignored in the computation of $\kappa$ \cite{Sachdev-1,Patel2}.
The energy current, whose correlation function determines the remaining
piece of the thermal conductivity, is given by 
\begin{equation}
\hat{j}_{E}=\int_{k}\sum_{i}\frac{k}{2m}\psi_{k,i}^{\dagger}\sigma_{x}\partial_{\tau}\psi_{k,i}.\label{jt}
\end{equation}
In the present particle-hole symmetric case, the thermal conductivity
is then given by the same diagrams as in the case of $\sigma$. Following
the same prescription as above,
\begin{equation}
\kappa=\mathrm{lim}_{\omega\rightarrow0}\frac{\mathrm{\mathcal{K}}_{E}^{\text{ret}}(\omega)}{\omega T},\label{eq:kappa}
\end{equation}
where $\mathrm{\mathcal{K}}_{E}(\tau)=\langle T[\hat{j}_{E}(0,x)\hat{j}_{E}(\tau,x)\rangle$
is a thermal current-current correlation function. This leads to
\begin{equation}
\kappa=\frac{Ne^{2}}{\sqrt{2m}J^{\frac{2}{5}}}T^{\frac{3}{5}}I_{2},
\end{equation}
where
\begin{equation}
I_{2}=\frac{1}{\pi^{2}}\int_{0}^{\infty}\text{d}z\,z^{2}\frac{\text{Im}\left[F_{\Sigma}(z)\right]^{2}}{\cosh^{2}(\frac{z}{2})}\int_{0}^{\infty}\text{d}y\frac{y^{2}}{\left|y^{2}+F_{\Sigma}\left(z\right)\right|^{4}}.\label{I2}
\end{equation}

In order to calculate integrals $I_{1}$ and $I_{2}$, one needs to
perform a numerical analytical continuation of the scaling functions
obtained in section III. In the spirit of SYK models, we compute these
integrals assuming the scaling form to be valid over the entire range
of frequencies. Numerical analytical continuation is a challenging
problem. The numerical integrals were done with the Pade approximation
method in the TRIQS library \cite{Serene,TRIQS}. 

We numerically find that slight variations in the Matsubara scaling
forms can significantly affect the integrals $I_{1}$ and $I_{2}$,
but their ratio is insensitive to numerical issues with the analytical
continuation process in the regime of interest. The Lorentz ratio
is
\begin{equation}
L=\frac{\kappa}{T\sigma}=\frac{I_{2}}{I_{1}}\approx3.2,\label{L}
\end{equation}
 which is rather close to the Fermi liquid value of $L_{FL}=\pi^{2}/3.$

\section{Discussion}

In this work, we studied a simple semimetallic version of a dispersive
SYK model in one dimension. Contrary to most studies of dispersive
models in the literature \cite{Zhang,Balents,Sachdev-1,Senthil,Shenoy},
we focus on the strongly dispersive limit, which corresponds to the
stable fixed point of this problem from the scaling point of view.
In this limit, we find that the Schwinger-Dyson equations do not admit
an exact analytic finite temperature solution even in the infrared
approximation, where it is assumed that $\Sigma(i\omega)\gg i\omega$.
In particular, the model does not exhibit conformal symmetry, which
makes it difficult to solve the Schwinger-Dyson equations analytically.
We solve those equations \emph{exactly} exploiting the scaling symmetry
of the model, combined with numerical calculations. We find that the
Greens function and self-energy scale with temperature with a power
law of $T^{-\frac{2}{5}}$ and $T^{\frac{4}{5}}$ respectively.

Using this solution to study transport properties, we show that dc
resistivity scales with a sublinear power law dependence on temperature,
$\rho\sim T^{\frac{2}{5}}.$ We compute the Lorentz ratio $L=\kappa/(\sigma T)$
with the analytically continued scaling functions and find that $L\approx3.2,$
rather close to that of Fermi liquids. The scaling analysis of this
problem indicates that if one starts in the high temperature SYK fixed
point of the problem, where $t^{2}/J\ll T\ll J$, the hopping parameter
will grow as one scales the temperature down, while the SYK coupling
is marginal. The scaling flows towards the strongly dispersive regime,
where the Schwinger-Dyson equations indicate the presence of a distinct
incoherent NFL regime at $T\ll t^{2}/J$. That contrasts with the
behavior of incoherent metals, which have a finite Fermi surface.
In the latter, the system flows towards a Fermi liquid at low temperature
\cite{Senthil}.

Those results should be compared with the several lattice models of
SYK dots that have been studied recently \cite{Sachdev-1,Senthil,Shenoy}.
Ref. \cite{Sachdev-1,Senthil} studied a lattice of coupled dots in
the limit where the SYK coupling $J$ is the highest energy scale.
In those cases, the physics of a single dot dominates, with the effects
of hopping being perturbative. The $\Sigma\propto\sqrt{\omega}$ scaling
of the self energy in this limit ultimately leads to a linear in $T$
dc resistivity.

Among the dispersive SYK models, the one studied in Ref. \cite{Shenoy}
is the closest to ours. They examined a two-band model for arbitrary
dimension and dispersion, with a color site dependent SYK interaction,
which forces the saddle point solution of the Green's function to
be diagonal but still color site dependent. Their solution for the
self-energy is purely imaginary and color site independent, differently
from our results. That leads to an approximate conformal symmetry
in the problem in the NFL regime, in contrast with our work, where
we find that conformal symmetry is absent. In this paper, we focused
in the crossover between the regime dominated by the $0+1$ dimensional
SYK fixed point and the low temperature NFL regime for a 1D semimetal
with parabolic band touching, and addressed the transport properties
of this novel state.

\emph{Acknowledgements.$-$ }We thank A. Haldar for several helpful
discussions. BU and GJ acknowledge Carl T. Bush fellowship and NSF
grant DMR-2024864 for partial support.

\end{document}